\documentclass[12pt,preprint]{aastex}

\usepackage{emulateapj5}

\begin{document}

\title{Time-dependent evolution of quasi-spherical, self-gravitating accretion flow}

\author{Mohsen Shadmehri
\affil{Department of Physics, School of Science, Ferdowsi University,
Mashhad, Iran}}\email{mshadmehri@science1.um.ac.ir}

\begin{abstract}
A self-similar solution for time evolution of quasi-spherical,
self-gravitating accretion flows is obtained under the assumption
that the generated heat by viscosity is retained in the flow. The
solutions are parameterized by the ratio of the mass of the
accreting gas to the central object mass and the viscosity
coefficient. While the density and the pressure are obtained
simply by solving a set of ordinary differential equations, the
radial and the rotational velocities are presented analytically.
Profiles of the density and the rotational velocities show two
distinct features. Low density outer accreting flow with
relatively flat rotation velocity, surrounds an inner high
density region. In the inner part, the rotational velocity
increases from the center to a transition radius where separates
the inner and outer portions. We show that the behaviour of the
solutions in the inner region depends on the ratio of the heat
capacities, $\gamma$, and the viscosity coefficient, $\alpha$.
\end{abstract}


\section{Introduction}
Accretion processes  are now believed to play a major role in
many astrophysical objects, from protostars to disks around
compact stars and AGN. Such systems have been studied at different
levels depending on their physical properties. Geometry of the
disk (thin or thick), transport of the thermal energy inside the
disk, self-gravity of the accreting gas and the magnetic fields
are among the most important factors which shape any theory for
such systems. For simplicity, traditional models of accretion
disks assume geometrically thin configuration and neglect
self-gravity of the accreting material (Shakura and Sunyaev 1973;
Pringle 1981). These models have been extended by considering
large scale magnetic fields, polytropic equation of state and
better understanding of the mechanism of angular momentum
transport (e.g., Livio and Pringle 1992; Ogilvie 1997; Duschl,
Strittmatter and Biermann 2000). However, the key ingredient in
all such models is that the generated heat by turbulent viscosity
does not remain in the flow. In other words, all of the viscously
dissipated energy was assumed to be radiated away immediately.

Another type of accretion flow known as Advection-Dominated
Accretion Flow (ADAF) has been proposed, in which generated heat
by viscosity can not escape from the system and retains  in the
flow (Ichimaru 1977; Narayan and Yi 1994). During recent years,
ADAFs have been paid attention as plausible states of accretion
flows around black holes, active galactic nuclei or dim galactic
nuclei (for review see Kato, Fukue and Mineshige 1998). Using
similarity method, Ogilvie (1999) (hereafter; OG) extended
original steady state self-similar solutions to time dependent
case. His solutions describe quasi-spherical time dependent
advection dominated accretion flows.

On the other hand, many authors tried to study the effects
related to the disk self-gravity. Recent observations show
significant deviations from Keplerian rotation in objects that
are believed to be accretion disk,  and in some cases, there are
strong evidences that the amount of mass in the disk is large
(e.g., Drimmel 1996; Greenhill 1996). The role of self-gravity in
accretion disks was discussed by Paczy\'{n}ski (1978), who studied
vertical structure of the disk under the influence of
self-gravity. Some authors investigated role of self-gravity on
waves in the disk (Lin and Pringle 1987; Lin and Pringle 1990).
Within the framework of geometrically thin configuration,
Mineshige and Umemura (1996) extended classical self-similar ADAF
solution (Narayan and Yi 1994) and found a global one dimensional
disk solutions influenced by self-gravity both in the radial and
perpendicular directions to the disk. In another study, Mineshige
and Umemura (1997) extended previous steady solutions to the time
dependent case while the effect of self-gravity of the disk had
been taken into account. They used isothermal equation of state
and so their solutions describe a viscous accretion disk in slow
accretion limit. Then, Mineshige, Nakayma and Umemura (1997)
extended this study by obtaining solutions for  polytropic
viscous accretion disks. Also, Tsuribe (1999) studied
self-similar collapse of an isothermal viscous accretion disk.

Although solution of Mineshige and Umemura (1996) describes
self-gravitating ADAFs, applicability of the solution restricts
to geometrically thin configurations. The most appropriate
geometrical configuration for advective flows, rather than thin,
is quasi-spherical as has been confirmed by many authors (e.g,
Narayan and Yi 1995). OG presented the first semi-analytical
quasi-spherical, time-dependent ADAFs solution. Here we note that
although OG considered advection-dominated accretion flow in a
point mass potential, it is straightforward to relax this
assumption so as to describe the effect of self-gravitation on
quasi-spherical advection-dominated accretion flows using
time-dependent similarity solutions.

This paper is organized as follows. In section 2 the general
problem of constructing a model for quasi-spherical,
self-gravitating accretion flow is defined. The self-similar
solutions are presented in section 3, and the effects of the input
parameters are examined. We summarize  the results in section 4.

\section{Formulation of the Problem}
We start with the approach adopted by Ogilvie (1999) who studied
quasi-spherical accretion flow without self-gravity. In his
approach, the equations are written in spherical coordinates $(r,
\theta, \varphi )$ by considering the equatorial plane
$\theta=\frac{\pi}{2}$ and neglecting terms with any $\theta$ and
$\varphi$ dependence. It implies  that each physical variable
represents approximately spherically averaged quantity. So, all
the physical quantities depend only on the spherical radius $r$
and time $t$. Self-gravity of the accreting gas can be considered
simply by Poisson equation and the corresponding term in the
momentum equation. The governing equations are the continuity
Equation,
\begin{equation}
\frac{\partial\rho}{\partial t} +
\frac{1}{r^2}\frac{\partial}{\partial r}(r^2 \rho v_{\rm
r})=0,{\label{eq:con}}
\end{equation}
the Equations of motion,
\begin{equation}
\frac{\partial v_{\rm r}}{\partial t} + v_{\rm r}\frac{\partial
v_{\rm r} }{\partial r} +\frac{1}{\rho}\frac{\partial p}{\partial
r } + \frac{\partial \Psi}{\partial r} +\frac{GM_{\star}}{r^2}
=\frac{v_{\rm\varphi}^2}{r},
\end{equation}
\begin{equation}
\rho(\frac{\partial}{\partial t}(rv_{\rm\varphi})+v_{\rm
r}\frac{\partial}{\partial
r}(rv_{\rm\varphi}))=\frac{1}{r^2}\frac{\partial}{\partial r}[\nu
\rho r^4 \frac{\partial}{\partial r}(\frac{v_{\rm\varphi}}{r})],
\end{equation}
the Poisson's Equation,
\begin{equation}
\frac{1}{r^2}\frac{\partial}{\partial r}(r^2
\frac{\partial\Psi}{\partial r})=4\pi G\rho,
\end{equation}
and the energy Equation,
\begin{equation}
\frac{1}{\gamma - 1}(\frac{\partial p}{\partial t}+ v_{\rm
r}\frac{\partial p}{\partial r})+
\frac{\gamma}{\gamma-1}\frac{p}{r^2}\frac{\partial}{\partial
r}(r^2 v_{\rm r})=\nu\rho r^{2}[\frac{\partial}{\partial
r}(\frac{v_{\rm\varphi}}{r})]^2.
\end{equation}

In order to solve the equations, we need to assign the kinematic
coefficient of viscosity $\nu$. Although there are many
uncertainties about the exact form of viscosity, authors
introduce some prescriptions for $\nu $ regarding dimensional
analysis or just based on phenomenological considerations (e.g,
Duschl, Strittmatter and Biermann 2000). We employ the usual
$\alpha$ prescription for the viscosity, which we write as
(Shakura and Sunyaev 1973)
\begin{equation}
\nu=\alpha\frac{p}{\rho \Omega_{\rm k}},
\end{equation}
where $\Omega_{\rm k}=(GM_{\star}/r^{3})^{1/2}$ is the Keplerian
angular velocity at radius $r$. This prescription originally
introduced for viscosity in thin accretion disk, however, it has
been widely used for studying dynamics of thick accretion disk,
such as ADAFs.

 To simplify the equations, we make the following substitutions:
\begin{equation}
\rho \rightarrow \hat{\rho} \rho, p \rightarrow \hat{p} p,
v_{r,\varphi} \rightarrow \hat{v} v_{r,\varphi}, \Psi\rightarrow
\hat{\Psi} \Psi, r \rightarrow \hat{r} r, t \rightarrow \hat{t} t,
\end{equation}
where
\begin{equation}
\hat{v}=\sqrt{\frac{GM_{\star}}{\hat{r}}}=\frac{\hat{r}}{\hat{t}},
\hat{t}=\frac{1}{\sqrt{4\pi G\hat{\rho}}},\hat{p}=\hat{\rho}
\hat{v}^{2}, \hat{\Psi}=\frac{GM_{\star}}{\hat{r}}.
\end{equation}
Under these transformation, the continuity equation does not
change and the rest of equations are cast into
\begin{equation}
\frac{\partial v_{\rm r}}{\partial t} + v_{\rm r}\frac{\partial
v_{\rm r} }{\partial r} +\frac{1}{\rho}\frac{\partial p}{\partial
r } + \frac{\partial \Psi}{\partial r} +\frac{1}{r^2}
=\frac{v_{\rm\varphi}^2}{r},{\label{eq:mom}}
\end{equation}
\begin{equation}
\rho(\frac{\partial}{\partial t}(rv_{\rm\varphi})+v_{\rm
r}\frac{\partial}{\partial
r}(rv_{\rm\varphi}))=\frac{\alpha}{r^2}\frac{\partial}{\partial
r}[p r^{11/2} \frac{\partial}{\partial
r}(\frac{v_{\rm\varphi}}{r})],
\end{equation}
\begin{equation}
\frac{1}{r^2}\frac{\partial}{\partial r}(r^2
\frac{\partial\Psi}{\partial r})=\rho,{\label{eq:pos}}
\end{equation}
\begin{equation}
\frac{1}{\gamma - 1}(\frac{\partial p}{\partial t}+ v_{\rm
r}\frac{\partial p}{\partial r})+
\frac{\gamma}{\gamma-1}\frac{p}{r^2}\frac{\partial}{\partial
r}(r^2 v_{\rm r})=\alpha p r^{7/2}[\frac{\partial}{\partial
r}(\frac{v_{\rm\varphi}}{r})]^2.{\label{eq:ener}}
\end{equation}

\section{Self-Similar Solutions}

We look for self-similar solutions of Equations (\ref{eq:con}) and
(\ref{eq:mom})-(\ref{eq:ener}) and reduce this system into a set
of ordinary differential equations. We define a self-similar
variable
\begin{equation}
\xi=\frac{r}{(t_0-t)^{2/3}},
\end{equation}
where $t<t_{\rm 0}$ and demand that
\begin{equation}
\rho (r, t) = (t_0 - t)^{-2} R(\xi),
\end{equation}
\begin{equation}
\ p(r, t) = (t_0 - t)^{-8/3} P(\xi),
\end{equation}
\begin{equation}
\ v_{\rm r}(r, t) = (t_0 - t)^{-1/3} V(\xi),\label{eq:vr}
\end{equation}
\begin{equation}
\ v_{\rm\varphi}(r, t) = (t_0 - t)^{-1/3} \Phi(\xi),\label{eq:vp}
\end{equation}
\begin{equation}
\Psi(r, t) = (t_0 - t)^{-2/3} S(\xi).
\end{equation}
Substituting these expressions into the above Equations, we
obtain the following set of self-similar Equations:
\begin{equation}
\ -\frac{2}{9}\xi +\frac{1}{R}\frac{dP}{d\xi}+
\frac{dS}{d\xi}+\frac{1}{\xi^2}=\frac{\Phi^2}{\xi},\label{eq:motion}
\end{equation}
\begin{equation}
\ -\xi^3 R (\Phi + 2\xi \frac{d\Phi}{d\xi})=3\alpha
\frac{d}{d\xi}[P
\xi^{11/2}\frac{d}{d\xi}(\frac{\Phi}{\xi})],\label{eq:ang}
\end{equation}
\begin{equation}
\frac{1}{\xi^2}\frac{d}{d\xi}(\xi^2\frac{dS}{d\xi})=R,\label{eq:poss}
\end{equation}
\begin{equation}
\frac{2}{3}(\frac{4-3\gamma}{\gamma-1})=\alpha \xi^{7/2}
[\frac{d}{d\xi}(\frac{\Phi}{\xi})]^2.{\label{eq:ener1}}
\end{equation}

Interestingly, the continuity Equation is integrable and gives
$\xi^{2} R (\xi+3V/2)=C$, where $C$ is a constant and should be
zero. Thus,
\begin{equation}
\ V=-\frac{2}{3} \xi.
\end{equation}
Also, we can obtain rotational similarity function $\Phi(\xi)$
simply by integrating Equation (\ref{eq:ener1}),
\begin{equation}
\Phi(\xi)=\frac{\xi}{\xi_{\rm
s}}\Phi_{s}+\frac{4\lambda}{3}(\xi^{1/4}-\xi_{\rm
s}^{-3/4}\xi).{\label{eq:phi}}
\end{equation}
where
\begin{equation}
\lambda=\sqrt{\frac{2}{3\alpha}(\frac{4-3\gamma}{1-\gamma})},
\end{equation}
and $\Phi_{\rm s}$ is rotational velocity at some $\xi_{\rm s}$.
For simplicity, we assume that $\xi_{\rm s}$ defines the outer
boundary of the accretiong gas. This Equation correctly gives
$\Phi(\xi=0)=0$, and reaches to a maximum $\Phi_{\rm m}$ by
increasing $\xi$ from zero to a point at $\xi_{\rm m}$. One can
simply show
\begin{equation}
\xi_{\rm m}=\frac{1}{4\sqrt[3]{4}}(\frac{\xi_{\rm s
}}{\sqrt[4]{\xi_{\rm s}}-\sqrt[4]{\xi_{\rm 0}}})^{4/3},
\end{equation}
where $\xi_{\rm 0}=3\Phi_{\rm s}/4\lambda$ and
\begin{equation}
\Phi_{\rm m}=\frac{\lambda}{\sqrt[3]{4}}(\frac{\xi_{\rm s
}}{\sqrt[4]{\xi_{\rm s}}-\sqrt[4]{\xi_{\rm 0}}})^{1/3}.
\end{equation}
Clearly, behaviour of rotational velocity and the other physical
variables except for the radial velocity, depend on  $\xi_{\rm s}$
and $\Phi_{\rm s}$. We can consider both $\xi_{\rm s}$ and
$\Phi_{\rm s}$ as free parameters, however, it's possible to
determine them uniquely, if we parameterize self-similar solutions
using conserved quantities, e.g. mass of the system. As can be
seen from the definition of $\lambda$, the above self-similar
solutions are applicable within the range of $1<\gamma <4/3$ if
we consider a positive value for $\alpha$.

We can derive asymptotic solutions when approaching the origin
$\xi=0$ as follows
\begin{equation}
\Phi(\xi) \longrightarrow \frac{4}{3}\lambda
\xi^{1/4},\label{eq:asy1}
\end{equation}
\begin{equation}
\ R(\xi) \longrightarrow
4(\frac{2}{3}\lambda^{2}+\frac{1}{11\alpha})
\xi^{-3/2},\label{eq:asy2}
\end{equation}
\begin{equation}
\ P(\xi) \longrightarrow
\frac{32}{33\alpha}(\frac{2}{3}\lambda^{2}+\frac{1}{11\alpha})
\xi^{-1}.\label{eq:asy3}
\end{equation}
These asymptotic solutions are valuable when performing numerical
integrations to obtain similarity solutions starting from
$x\rightarrow 0^{+}$. Note that the behaviour of the solutions
near to origin only depends on $\lambda$ and $\alpha$.

 We now proceed to solve the rest of above similarity Equations.
 By substituting Equation (\ref{eq:phi}) into Equation
 (\ref{eq:ang}), we obtain
\begin{equation}
\frac{15}{4}P+\xi\frac{dP}{d\xi}=f(\xi) R.\label{eq:main1}
\end{equation}
where
\begin{equation}
f(\xi)=\frac{\xi^{1/4}}{3\alpha\lambda}[3\frac{\Phi_{\rm
s}}{\xi_{\rm s}}\xi+4\lambda (\frac{1}{2}\xi^{1/4}-\xi_{\rm
s}^{-3/4}\xi)].
\end{equation}
After some algebra, from this Equation and Equations
(\ref{eq:motion}) and (\ref{eq:poss}), the following Equation is
obatined:
\begin{equation}
\frac{15}{4}f P \frac{d^{2}P}{d\xi^{2}}=S_{1}(\xi, P,
dP/d\xi)+S_{2}(\xi, P, dP/d\xi)+S_{3}(\xi, P,
dP/d\xi),\label{eq:main}
\end{equation}
where
\begin{equation}
S_{1}=\frac{19}{4}f(\frac{dP}{d\xi})^{2}-(\frac{df}{d\xi})
(\frac{dP}{d\xi}) (\frac{15}{4}P+\xi \frac{dP}{d\xi}),
\end{equation}
\begin{equation}
S_{2}=\frac{g}{\xi^{2}}(\frac{15}{4}P+\xi\frac{dP}{d\xi})^{2}-\frac{1}{f}(\frac{15}{4}P+\xi\frac{dP}{d\xi})^{3},
\end{equation}
\begin{equation}
S_{3}=-2\frac{f}{\xi}
\frac{dP}{d\xi}(\frac{15}{4}P+\xi\frac{dP}{d\xi}),
\end{equation}
\begin{equation}
g(\xi)=\frac{d}{d\xi}(\frac{2}{9}\xi^{3}+\xi\Phi^{2}).
\end{equation}

This Equation can easily be integrated considering appropriate
boundary conditions. We can parameterize the solutions by
$\xi_{\rm s}$ as a function of the mass of the accreting gas and
the viscosity parameter $\alpha$. For simplicity, we assume
$R(\xi_{\rm s})=R_{\rm s }$ and $P(\xi_{\rm s})=P_{\rm s }$. Total
mass of the accreting gas is
\begin{equation}
M=4\pi \int_{0}^{\infty} \rho r^{2}dr.
\end{equation}
Using similarity solutions, this Equation reads
\begin{equation}
M=M_{\star}\int_{0}^{\xi_{\rm s}} R \xi^{2} d\xi.
\end{equation}
Substituting from  Poisson Equation (\ref{eq:poss}), we have
\begin{equation}
M=M_{\star} \xi_{\rm s}^{2} \frac{dS}{d\xi}|_{\xi_{\rm s}},
\end{equation}
By inserting $dS/d\xi$ from Equation (\ref{eq:motion}) into the
above Equation, we finally obtain
\begin{equation}
M=M_{\star} [\xi_{\rm s} \Phi_{\rm s}^{2}+\frac{2}{9}\xi_{\rm
s}^{3}-\frac{\xi_{\rm s}^2}{R_{\rm s}}(\frac{dP}{d\xi})_{\rm
s}-1].\label{eq:set1}
\end{equation}
Another relation is obtained from Equation (\ref{eq:main1}), as
follows
\begin{equation}
\frac{15}{4}P_{\rm s}+\xi_{\rm s}(\frac{dP}{d\xi})_{\rm
s}=f(\xi_{\rm s})R_{\rm s},\label{eq:set2}
\end{equation}
where
\begin{equation}
f(\xi_{\rm s})=\frac{\sqrt[4]{\xi_{\rm
s}}}{3\alpha\lambda}(3\Phi_{\rm s}-2\lambda \sqrt[4]{\xi_{\rm
s}}).
\end{equation}
Now, we can use the standard fourth-order Runge-Kutta scheme to
integrate nonlinear ordinary differential equation (\ref{eq:main})
from sufficiently small $\xi$ to $\xi_{\rm s}$.  Given $\xi_{\rm s
}$ and $\Phi_{\rm s}$ and using asymptotic solutions near to the
origin (i.e., equations (\ref{eq:asy2}) and (\ref{eq:asy3})), one
can start numerical integration and equation (\ref{eq:set1}) gives
corresponding ratio of masses, i.e. $M/M_{\star}$. Before
presenting  results of integration, it would be interesting to
explore the typical behaviour of rotational velocity $\Phi(\xi)$.

\begin{figure}
\plotone{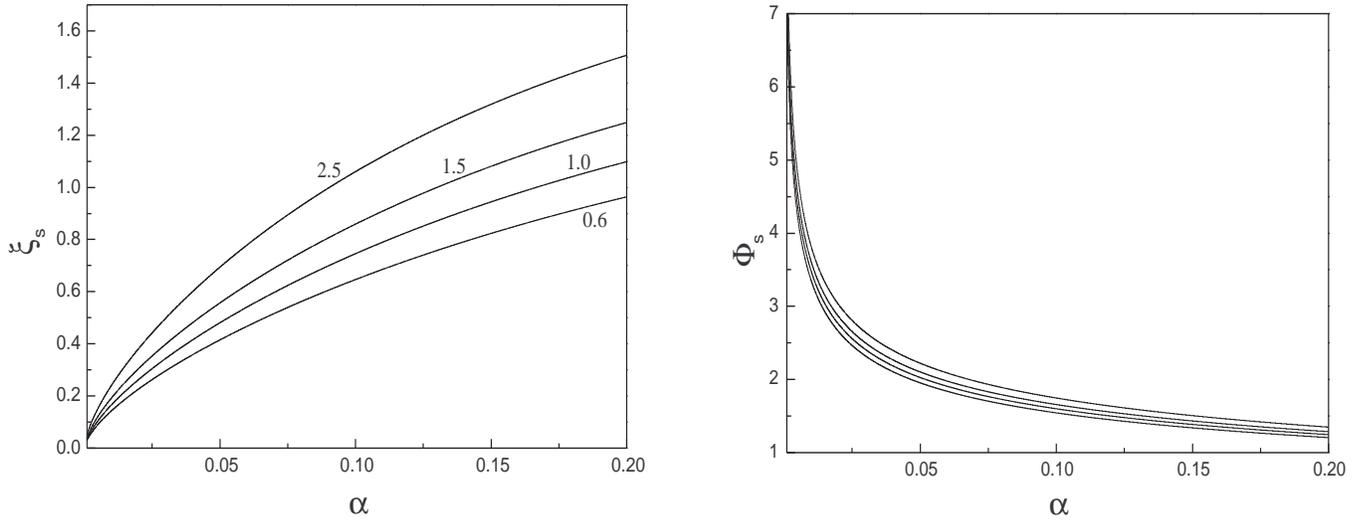} \caption{Typical behaviour of $\xi_{\rm s}$ and
$\Phi_{\rm s}$ as a function of $\alpha$ corresponding to
$\gamma=5/4$, $P_{\rm s}=0$ and $R_{\rm s}\simeq 0 $. Left: Each
curve is marked by the ratio of mass of the accreting gas to the
mass of the central object, $M/M_{\star}$. Right: While the
lowest curve corresponds to $M/M_{\star}=0.6$, the curve shifts
upward as this ratio increases. }\label{fig:fig1}
\end{figure}

Just as an illustrative example we follow another approach: lets
assume that we know $R_{\rm s}$, $P_{\rm s}$ and $M/M_{\star}$
from numerical integration and so, one can determine $\xi_{\rm
s}$ and $\Phi_{\rm s}$ from Equations (\ref{eq:set1}) and
(\ref{eq:set2}) uniquely. For example, if we set $P_{\rm s}=0$ and
$(dP/d\xi)_{\rm s}=0$ and consider non-zero but negligible
$R_{\rm s}$, the outer boundary of the system is determined
analytically
\begin{equation}
\xi_{\rm
s}=\sqrt[3]{2\lambda^{4}+4.5(\frac{M}{M_{\star}}+1)-\sqrt{4\lambda^{4}(\lambda^{4}+4.5(\frac{M}{M_{\star}}+1))}},
\end{equation}
and
\begin{equation}
\Phi_{\rm s}=\frac{2}{3}\lambda \sqrt[4]{\xi_{\rm s}}.
\end{equation}
In this case, Figure \ref{fig:fig1} shows $\xi_{\rm s}$ and
$\Phi_{\rm s}$ as functions of $\alpha$ for $\gamma=5/4$ and
$M/M_{\star}=0.6, 1, 1.5$ and $2.5$. As this Figure shows for
fixed ratio of masses, the outer radius of the system increases
by increasing the viscosity coefficient $\alpha$. Also, if the
ratio of masses increases, the outer radius increases
irrespective of the exact value of $\alpha$. However, rotational
velocity at $\xi_{\rm s}$ (i.e. $\Phi_{\rm s}$) is not very
sensitive to values of $\alpha$ or $M/M_{\star}$ for large values
of $\alpha$. If the viscosity coefficient $\alpha$ tends to small
values, on the other hand, the rotational velocity $\Phi_{\rm s}$
increases significantly.

\begin{figure}
\plotone{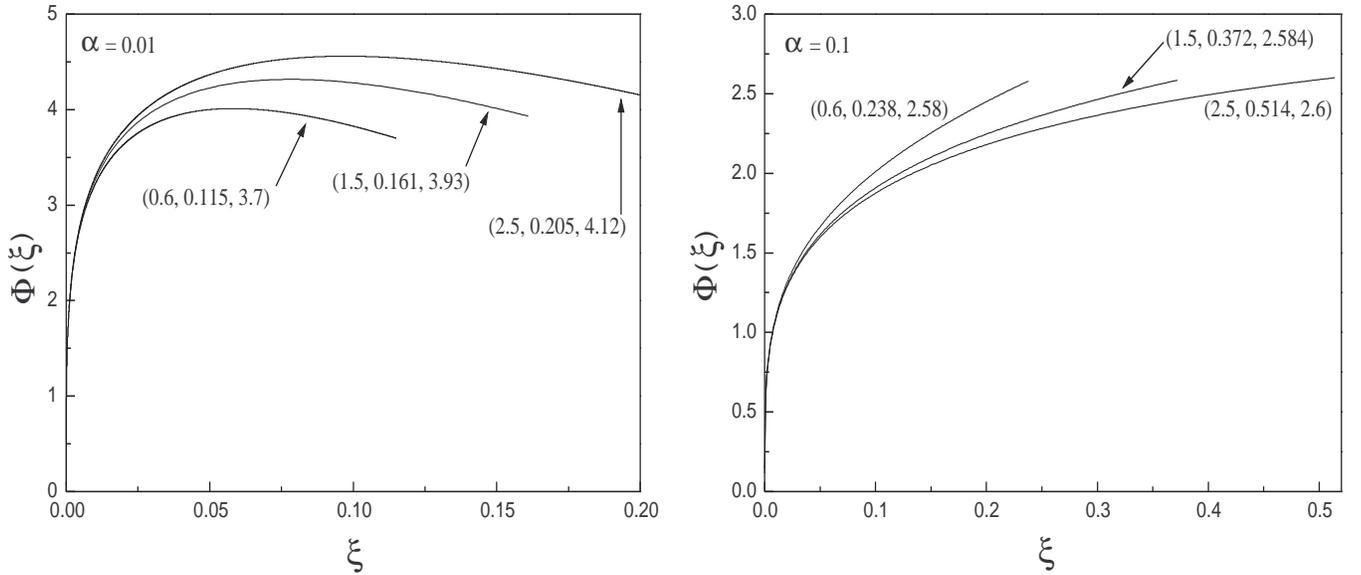} \caption{Self-similar rotational velocity
corresponding to $\gamma=5/4$ and $R_{\rm s}=P_{\rm s}=0.01$.
Each curve is marked by the ratio of the mass of accreting gas to
the mass of the central object, $M/M_{\star}$, and $\xi_{\rm s}$
and $\Phi_{\rm s}$ as $(M/M_{\star}, \xi_{\rm s}, \Phi_{\rm s})$.
}\label{fig:figure2}
\end{figure}

In  Figure \ref{fig:figure2}, we plot rotational velocity of
representative cases with $\alpha=0.01, 0.1$ and
$M/M_{\star}=0.6, 1.5, 2.5$. As this Figure shows, rotation of
the flow increases from the center to the outer radius. There are
two regimes as rotation of the flow concerns. While in the outer
part, profile of the rotational velocity in similarity space is
nearly flat, in the inner portion this velocity strongly
increases. However, for low values of $\alpha$ (e.g. $0.01$),
rotational velocity reaches to a maximum then with nearly
constant and small slope decreases. Also, the rotational velocity
increases as the parameter $\alpha$ decreases and the outer
region with flat rotation profile becomes larger, as the ratio of
the mass of the accreting gas to the  mass of the central object
increases. In other words, while rotational behaviour of the
inner part is nearly independent of the ratio of masses, the
outer portion's rotation and it's extension are sensitive to that
ratio.

\begin{figure}
\plotone{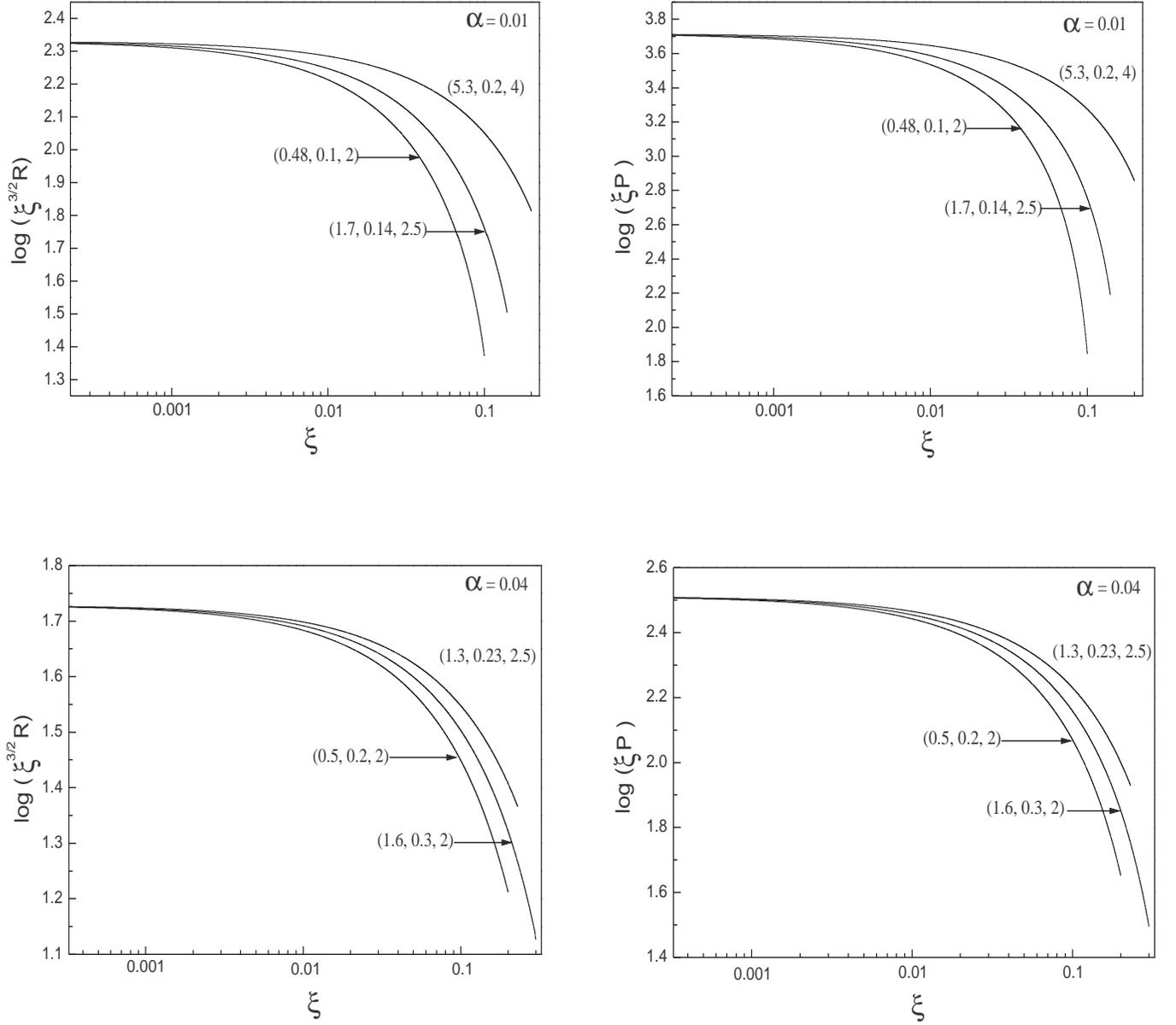} \caption{Self-similar density and the pressure
corresponding to $\gamma=5/4$  for different values of viscosity
coefficient, i.e. $\alpha=0.01$ and $0.04$. Curves are labeled the
same as Figure 2.}\label{fig:figure3}
\end{figure}

In Figure \ref{fig:figure3}, typical behaviour of the density and
the pressure in similarity space are shown. We can see in the
inner region the solutions can clearly be described by asymptotic
solutions (\ref{eq:asy2}) and (\ref{eq:asy3}), however, outer
region has different density and pressure profiles.
Interestingly, behaviour of solutions in the inner region is
independent of the extension of the system ($\xi_{\rm s}$) and
the rotational velocity $\Phi_{\rm s}$. As the ratio of
$M/M_{\star}$ increases, the extension of the outer region
becomes larger. Flows with large values of $\alpha$ have lower
central mass concentration comparing to accretions with low
values of $\alpha$. Generally, one can say quasi-spherical,
self-gravitating accretion flows of our model consist of two
parts, one inner portion with high density and an outer part with
nearly lower density and flat rotation profile. However, in both
regions the radial velocity is the same, irrespective of the
input parameters.

\begin{figure}
\plotone{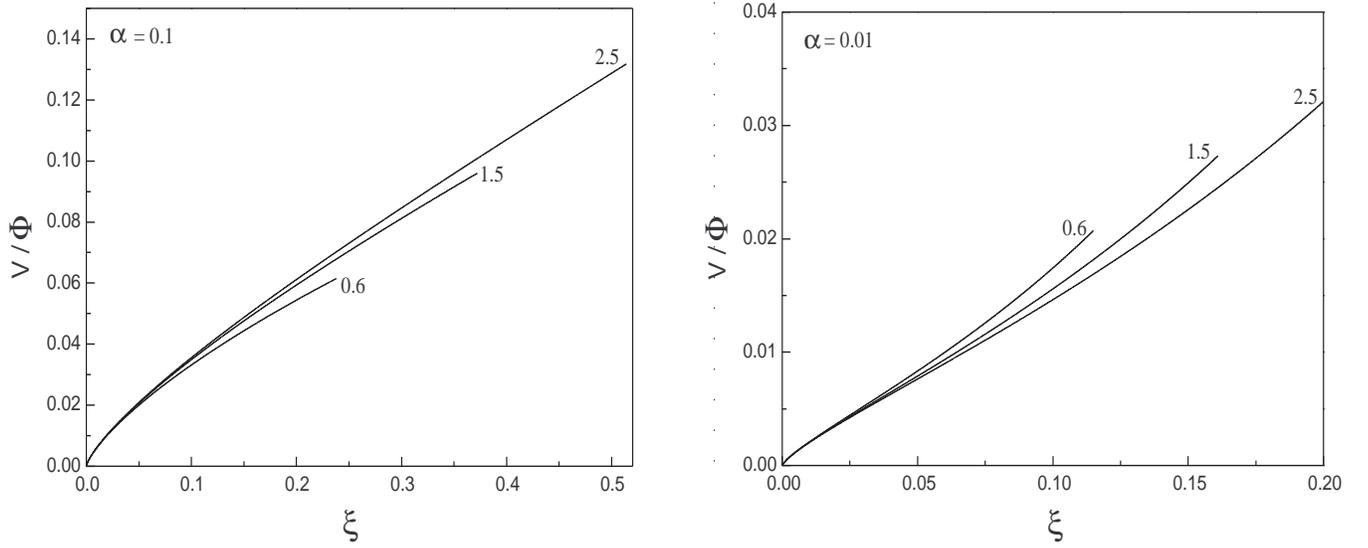} \caption{Ratio of the radial velocity to the
rotational velocity for $\gamma=5/4$ and $R_{\rm s}=P_{\rm
s}=0.01$ and $\alpha=0.1, 0.01$. Curves are labeled by the ratio
of the masses, i.e. $M/M_{\star}$ }\label{fig:figure4}
\end{figure}

From Equations (\ref{eq:vr}) and (\ref{eq:vp}), we can simply show
that the ratio of the radial velocity to the rotational velocity
is $v_{\rm r}(r,t)/v_{\rm\varphi}(r,t)=V/\Phi$. Figure
\ref{fig:figure4} shows this ratio for $\alpha=0.1$ and $0.01$
and three different values of $M/M_{\star}$. Clearly, this ratio
of velocities are much smaller than unity, implying that at each
radius $r$ and at any time $t$, rotational velocity is greater
than the radial velocity. However, as the viscosity parameter
$\alpha$ increases, the ratio increases as well. This behaviour
is easy to understand;  because we showed that radial similarity
velocity $V$ is proportional to similarity variable $\xi$,
independent of the input parameters. However, rotational
similarity velocity $\Phi$ decreases by increasing the parameter
$\alpha$, as can be seen from Figure \ref{fig:figure2}. So, the
ratio $V/\Phi$ increases in the case that $\alpha$ increases.

As we showed the total mass of the accreting gas is conserved. We
can simply write the integral representing the angular momentum
 as $J=4\pi \int\rho v_{\rm\varphi} r^3 dr$. Using the similarity
 solutions, we can show that the total angular momentum $J$ is
 proportional to $(t_0-t)^{1/3}$. In non-self-gravitating
 quasi-spherical accretion flow, $J$ is conserved and the total
 mass $M$ is proportional to $(t_0-t)^{-1/3}$ as has been shown by OG. But our solutions
 imply that the total mass of accreting material is conserved  and
angular momentum is decreasing. Also, the central density
$\rho_{\rm c}$ increases as $\rho_{\rm c}\propto (t_0-t)^{-2}$.
For non-self-gravitating flow, we see another behaviour:
$\rho_{\rm c}\propto (t_0-t)^{-7/3}$.  However, the radius of the
flow in both cases decreases in proportion to $(t_0-t)^{2/3}$.

\section{Summary}
In this paper, we have studied quasi-spherical accretion flow, in
which heat generated by viscosity retained in the flow. In
opposition to the usual studies performed up to now, we have
considered self-gravity of the flow. We derived similarity
solutions for such flows which are applicable within the range
$0<\gamma<4/3$, if we consider positive value of $\alpha$. Radial
and rotational velocities have been obtained analytically.
Obtained solutions  parameterized by the ratio of the disk mass
to the central object mass, $M/M_{\star}$, and the viscosity
parameter, $\alpha$. We showed that the extension of the
accreting gas depends on this ratio and the viscosity parameter.
More importantly, these input parameters have direct effect on the
rotational velocity.

Our solutions are different from the solutions by OG in various
respects. We found that the radial similarity velocity is in
proportion to $\xi$, implying no critical point. This fortunate
circumstance let us to integrate rest of the Equations simply,
although physically one should bear in mind this kind of velocity
(i.e., independent of input parameters) is as a result of
mathematical limitations of similarity method.  At the outer edge
of the accreting gas, the density and the pressure have low
values comparing to the central region.

Other viscosity laws have been proposed, for instance the
$\beta$-prescription which is based on analogy with turbulence
observed in laboratory sheared flows and gives $\nu \propto
v_{\rm\varphi} r$ (Duschl, Strittmatter and Biermann 2000).
Although we have not explored self-gravitating accretion with
this prescription, there are self-similar solutions with
similarity indices the same as those have been found with
$\alpha$-prescription in this paper. However, ordinary
differential equations governing the similarity physical
variables are different and should be solved numerically. It
would be interesting one compare the similarity solutions with
$\beta$-prescription with those we have obtained here.

\acknowledgements

I thank the anonymous referee for his/her report  which showed an
error  in the initial version of the paper, prompting  me to find
the correct form of the asymptotic solutions near to the origin.

\clearpage


\end{document}